\documentclass[journal]{IEEEtran}
\usepackage{graphicx}
\usepackage[fleqn]{amsmath}
\usepackage{authblk}
\usepackage{amsmath}

\ifCLASSINFOpdf
\else
\fi
\begin{document}
%
\title{Time Optimal Spectrum Sensing}
%
%
%

\author[1]{Garimella Rama Murthy}
\author[2]{Rhishi Pratap Singh}
\author[3]{Samdarshi Abhijeet}
\author[4]{Sachin Chaudhary}
\affil[1,2,3,4]{Signal Processing and Communication Research Center

International Institute of Information Technology, Hyderabad, India  }
\affil[1]{Email: rammurthy@iiit.ac.in}
\affil[2]{Email: rhishi.pratap@research.iiit.ac.in }
\affil[3]{Email: samdarshi.abhijeet@research.iiit.ac.in }
\affil[4]{Email: sachin.c@iiit.ac.in }

\maketitle

\begin{abstract}
Spectrum sensing is a fundamental operation in cognitive radio environment. It gives information about spectrum availability by scanning the bands. Usually a fixed amount of time is given to scan individual bands. Most of the times, historical information about the traffic in the spectrum bands is not used. But this information gives the idea, how busy a specific band is. Therefore, instead of scanning a band for a fixed amount of time, more time can be given to less occupied bands and less time to heavily occupied ones. In this paper we have formulated the time assignment problem as integer linear programming and source coding problems. The time assignment problem is solved using the associated stochastic optimization problem.  
\end{abstract}


%
\IEEEpeerreviewmaketitle

\section{Introduction}
%
%
%
%
In recent years, Cognitive Radio technology \cite{mitola} is proposed for making efficient utilization of electromagnetic spectrum. At the physical layer of cognitive radio networks, various techniques are proposed for Spectrum  Sensing \cite{yucek}. One of the basic approaches to spectrum sensing is based on Energy Detection. In earlier efforts of spectrum sensing, the temporal record/history of spectrum utilization has been completely ignored. Some researchers realized that such approach to spectrum sensing is sub-optimal\cite{sumit}. The authors particularly proposed “Doubly Cognitive Network Architecture” in which Intelligent Spectrum Sensing is carried out by taking the historical data of spectrum utilization into account. In this research paper, we make precise mathematical formulation of time optimal spectrum sensing and propose an interesting solution.


\section{Time Optimal Spectrum Sensing: Integer Linear programming}
Consider a band of EM spectrum available for wireless communication. Let this band be subdivided into sub-bands labeled ${1,2,\ldots,M}$. In traditional spectrum sensing based on, say, energy detection, all the sub-bands are scanned with a fixed, constant time irrespective of the historical data about packet traffic. It is logically clear that the sub bands which are heavily occupied(based on historical traffic data) can be scanned faster(sensing time is chosen to be smaller) while the less occupied sub-bands can be scanned using larger sensing time. The total available time for spectrum sensing of the entire band is assumed to be constant, say L seconds. The sensing time allocated for each of the sub bands is assumed to be integer valued. \\
\textbf{Note:} The time optimal spectrum sensing problem formulated below does not depend on the spectrum sensing approach.\\
\\
\textbf{Joint Detection-Estimation Approach to Spectrum Sensing:} In the following discussion we formulate the problem of prediction of packet traffic based on historical data as a Linear Mean Square Estimation problem. Also as in traditional spectrum sensing primary user detection is formulated as hypothesis testing based detection problem.\\
\\
As discussed earlier, we take the historical traffic data on various sub-bands into account for choosing the spectrum sensing time. In this direction we model the historical traffic data as an Auto Regressive(AR) process. In time, the unit on which the data is collected, could be an hour, day, month etc. Specifically, we fit a $p^{th}$ order AR process to the traffic data, i.e.
\begin{multline}
x(n+p) = a_1 x(n)+a_2 x(n+1)+\ldots\\ +a_p x(n+p-1) + w(n+p)
\end{multline}
where using LMSE(Linear Mean Square Error estimation i.e. Solving Yule-Walker equations) method, the coefficients are estimated and the traffic data is predicted(on certain time unit).\\
\textbf{Note:} The prediction tool(model) can be chosen to be more sophisticated (artificial Neural network based approach).\\
\\
Let the predicted data in “M” sub-bands be denoted by $n_1, n_2,\ldots, n_M$.
We normalize the number of packets in various sub-bands in the following manner
\begin{equation}
q_i = \frac{n_i} {\sum_{j=1}^{M} n_j}  \text{ for }\ 1 \leq i  \leq M
\end{equation}
Thus $\{q_1, q_2,\ldots, q_M\}$ is a probability mass function, associated with packet traffic data in various sub-bands.\\
\\
Now, we formulate the time-optimal spectrum sensing problem . Our goal is to allocate the total time for sensing the entire band ( say L seconds ) into time for sensing sub-	bands( i.e.$T_1, T_2,\ldots,T_M$) such that the average sensing time i.e.
\begin{equation}
\bar{T} = {\sum_{i=1}^{M} T_i q_i}  \text{  with  } {\sum_{i=1}^{M} T_i} = L 
\end{equation}
is minimized. We reason below that if no constraints are imposed on $\{T_i\}$, then we have a trivial problem.\\
\textbf{Case1:}
In this case order $\{q_i\}$ from smallest value to largest value, i.e. label them as 
$\{\hat{q_1}, \hat{q_2},\ldots,\hat{q_M}\}$
Set $T_1=L, T_2=0,\ldots, T_M=0$. With such a trivial allocation, $\bar{T}$ is minimized.\\
\\
\textbf{Case 2:} Minimum sensing time in any of the bands is lower bounded by $T_1$(i.e smallest sensing time is at-least $T_1$).
Allocation can be as following: $T_1,T_1,\ldots,T_1,(L-(M-1)T_1)$ with $(L-MT_1+T_1) \geq T_1$ \\
\\
\textbf{Case 3:} Smallest sensing time is at-least $T_1$ and other sensing times differ by at-least 1 time unit.
Allocation can be as following: $T_1,T_1+1,T_1+2,\ldots,T_1+M-2, (L-S))$ with $(L-S) \geq T_1$, where S=($T_1)+(T_1+1) + (T_1+2) +\ldots,(T_1+M-2)$. \\
\\
\textbf{Case 4:} Smallest sensing time is at-least $T_1$ and other sensing times differ by at-least d time units.
Allocation can be as following: $T_1,T_1+d,T_1+2d,\ldots,T_1+(M-2)d, (L-S))$ with $(L-S) \geq T_1$, where S=($T_1)+(T_1+1) + (T_1+2) +\ldots,(T_1+M-2)$.\\ 
\\
Thus we are naturally led to imposition of realistic (practical) constraints on the integer valued $T_i$'s.\\
\textbf{Case A:} $T_i$'s are in arithmetic progression. i.e. $T_1, T_1+d,\ldots, T_1+(M-1)d$. 
These times must add up to total sensing time, L. Thus, we have
\begin{equation}\label{eq:1}
\begin{split}
MT_1 + \frac{dM(M-1)}{2} &= L \\
2MT_1 + dM(M-1) &= 2L 
\end{split}
\end{equation}
\textbf{Note:} In the above equation M, the number of sub-bands and 'L', the total sensing time are known. $T_1$, d are unknown variables. Since  $T_1$, d are always constrained to be integers, we have a linear Diophantine equation of the form 
$aT_1 + bd = 2L$ , 
where a=2M \text{ and } b=M(M-1)
There are standard techniques for solving such an algebraic equation.\\\\
\textbf{Case B:} $T_i$'s are in Geometric progression. 
i.e. $T_1, (T_1)(d),(T_1)(d^2),\ldots, (T_1)(d^{M-1})$. They must add up to total sensing time L.
\begin{equation}\label{eq:2}
\begin{split}
T_1+(T_1)(d)+(T_1)(d^2)+\ldots+ (T_1)(d^{M-1}) &= L \\
T_1(1+d+d^2+\ldots+(d^{M-1}) &= L\\ 
T_1 \frac{(d^{M}-1)}{d-1}  &= L
\end{split}
\end{equation}
As discussed earlier {M, L} are known and {$T_1$, d} are unknown. Thus we need to solve the following algebraic equation
\begin{equation}
T_1 d^{M}-Ld-(T_1-L)=0
\end{equation}
\textbf{Goal:} To solve the above algebraic equation for {$T_1$, d} suppose we assume that d = 2. Thus we have to decide '$T_1$' for $T_1(2^M -1)-L=0$ 
\\
Thus, for a given 'M'; {$T_1,(2^M-1)$} must be divisors of L. If not, no solution exists.
Suppose 'M' is such that  $2^M-1$ is a prime i.e. A Mersenne prime. If 'L' happens to be a prime number, no solution exists. (It should be noted that 'M' must necessarily be a prime for $2^M-1$ to be a Mersenne prime). Thus in this case for a solution to exist 'L' must be such that its prime  factorization contains the Mersenne prime $2^M-1$.
For a given M if L is divisible by $2^M-1$, we have
\begin{equation}
T_1 = \frac{L}{(2^M -1)}
\end{equation}
\textbf{Significance of this solution:}
Energy detection is facilitated by the use of FFT of certain length/size. Typically the FFT sizes are  power of 2. Thus, 'd'  can be chosen to be a power of 2, leading to explicit solution for $T_1$, i.e.
\begin{equation}\label{eq:3}
T_1 = L\frac{d-1}{(d^M -1)}
\end{equation}
\textbf{General Solution in Case B:
}Factoring 'L' gives all the possibilities for $T_1$. A short computation will give the desired solutions, if any.
\\\\
\textbf{Justification of AP/GP for sensing times:}
As the probabilities decrease, the increase in sensing times assume values in an AP i.e. the rate of increase of sensing times is linear, or Sensing times increase geometrically 	(implemented by an FFT of suitable frequency resolution.) e.g. a, 2a, 4a, 8a, 16a,$\ldots$
\\\\
\textbf{Case C:} $T_i$'s are in Arithmetico-geometric sequence. 
i.e. $T_1, (T_1+d)(r),(T_1+2d)(r^2),\ldots, [T_1+(M-1)d](r^{M-1})$. They must add up to total sensing time L.
\begin{equation}
\begin{split}
T_1+(T_1+d)(r)+(T_1+2d)(r^2)+\ldots+\\ (T_1+(M-1)d)(r^{M-1}) &= L \\ 
T_1[1+r+r^{2}+\ldots+r^{M-1}]+dr[1+\\2r+3r^2+\ldots+(M-1)r^{M-2}] &=L \\
T_1 \frac{(1-r^{M})}{(1-r)}+dr[\frac{(1-Mr^{M-1})}{(1-r)} + \frac{(r-r^M)}{(1-r)^2}] &= L\\
\end{split}
\end{equation}
If common difference is equal to common ratio i.e d = r
\begin{equation}
T_1 \frac{(1-d^{M})}{(1-d)}+d^2[\frac{(1-Md^{M-1})}{(1-d)} + \frac{(d-d^M)}{(1-d)^2}] = L
\end{equation}
Thus, the Diophantine equation whose solutions are of interest to us are given by above equations. Solutions must be feasible/Non-negative integer values of $\{T_1,d\}$.
\\
\textbf{Note:} It can easily be reasoned that if d=0, the above equation reduces to \eqref{eq:1} and if r=1, (by using L'Hospitals rule) the equation reduces to \eqref{eq:3}.

\section{Time Optimal Spectrum Sensing: Source Coding}
In this section we relate the problem of “Time Optimal Spectrum Sensing” to the source coding problem.
\begin{equation}
\begin{split}
\mbox{Let } S_i &= n_i  \mbox{  for}\ 1 \leq i  \leq M \\
\mbox{Compute }\hat{p_i} &= \frac{S_i}{\sum_{j=1}^{M}S_j}  \mbox{  for}\ 1 \leq i  \leq M \\
\mbox{Let }\hat{T} &= \sum_{j=1}^{M}\hat{T_j} \hat{p_j} 
\end{split}
\end{equation}
Let X be the random variable assuming values $\{\hat{T_1},\hat{T_2},\ldots,\hat{T_M}\}$ with probabilities $\{\hat{p_1},\hat{p_2},\ldots,\hat{p_M}\}$.
\\
Shannon Entropy of X is given by 
\begin{equation}
H(X) =- \sum_{j=1}^{M}\hat{p_j} log\hat{(p_j)} 
\end{equation}
\\
Suppose we require the spectrum sensing times in various sub-bands i.e. $\{\hat {T_i}\}_{i=1}^{M}$ to satisfy 	the Kraft inequality. i.e.
\begin{equation}
\sum_{i=1}^{M} 2^{-\hat{T_i}} \leq 1
\end{equation}
\\
Then we necessarily have the following lower bound on average sensing time i.e. $\hat{T} \geq H(X)$
, where H(X) is the entropy of the random variable assuming values $\{\hat {T_i}\}_{i=1}^{M}$ with the 	probabilities $\{\hat {q_i}\}_{i=1}^{M}$.
In this connection we have following interesting lemma. 
\\
\textbf{Lemma:} If the sensing times $\{\hat {T_i}\}_{i=1}^{M}$ are increasing “at-least” in an arithmetical progression with common difference “1” i.e.
$ \hat{T_2}=\hat{T_1}+1, \hat {T_3}=\hat{T_1}+2,\ldots,\hat{T_M}=\hat {T_1}+(M-1)$
then Kraft inequality is satisfied.
\\
\textbf{Proof:} Refer \cite{abhijeet}
\\
\textbf{Note:} It is immediate that if Kraft inequality is satisfied with D = 2 i.e.
\begin{equation}
\sum_{i=1}^{M}{2^{- \hat {T_i}}} \leq 1 \text{  then  } \sum_{i=1}^{M}{D_0^{- \hat {T_i}}} \leq 1 
\end{equation}
for any $D_0>2$. Also if $\{\hat {T_i}\}_{i=1}^{M}$ increases faster than Arithmetic progression with common difference “ONE” (i.e. AP with common difference strictly greater than one or geometric progression etc) then Kraft inequality is satisfied. 
\\
Using Huffman coding we determine the values $\{\hat {T_i}\}_{i=1}^{M}$. Suppose they i.e. $\{\hat {T_i}\}_{i=1}^{M}$must add up to 'L'. Then the values of $\{\hat {T_i}\}_{i=1}^{M}$ are scaled/ normalized such that
$\sum_{i=1}^{M}{\hat {T_i}} = 1. $

\section{Numerical Experiments}
We now consider case A in section 2. We invoke the following theorem on computing the solution of linear Diophantine Equation \cite{andrews}.
\\
\textbf{Theorem:}
The linear Diophantine equation ax + by = c has a solution if and only if $d|c$ (d divides c), where d is 	G.C.D.(a,b). Furthermore if $(x_0, y_0$) is a solution for this equation, then the set of solutions of the equations consist of all integer pairs (x, y), where	$x=x_0+t(b/d)$ and $y=y_0+t( a/d)$ ,where t=$\ldots$,-2,-1,0,1,2,$\ldots$
\\
\textbf{Note:} We can compute any one solution discussed in the above theorem using Euclidean (G.C.D. Computation) algorithm.
\\
\textbf{Q:} How do we select the required solution?
i.e. $\{T_1,d\}$ should be non-negative.
\\
\textbf{Q:} What if there are multiple solutions for $\{T_1,d\}$?
\\\\	
\textbf{Examples:} Case of linear Diophantine Equation
\\
\textbf{case 1:} 
		$aT_1 + bd = 2L$, where a=2M, b=M(M-1) and L = Total sensing time.
\\		
let M=10, L=100 
\begin{equation}
\begin{split}
20T_1 + 90d &= 200 \\
\mbox{GCD }(20,90) &= 10. \mbox{ (200 is divisible by 10.)} \\
T_1 &= 1+t(90/10) \\
d &= 2-t(20/10)  
\end{split}
\end{equation}
For t=$\ldots$.-2,-1,0,1,2,$\ldots$ there are multiple solutions but there is only one interesting solution is with t=0, 
$T_1$ = 1 and d=2
\\
\textbf{case 2:} if GCD(a,b) in $aT_1+bd = 2L$ is a.
let M=15,
\begin{equation}
\begin{split}
30T_1+210d &= 1800 \\
\mbox{ GCD } (30,210) &= 30. \mbox{ (1800 is divisible by 30.)} \\
T_1+7d &= 60 \\
\end{split}
\end{equation}
One solution can be d=8 and $T_1$ =4 
\begin{equation}
\begin{split}
T_1 &= 4+t(210/30) \\
d &= 8-t(30/30)
\end{split}
\end{equation}
For t= 0,1,2,$\ldots$,7 there are solutions. So there are multiple 	solutions.
\\
\textbf{Note:}The solution for $\{T_1,d\}$ in case A and B is always a matching pair. 
\\
\textbf{Problem:} Suppose the number of solutions i.e. $\{T_1,d\}$ in case A, case B is strictly more than One.
\\
\textbf{Goal:} We would like to arrive at solutions that minimize both the mean and variance of sensing time random variable. Suppose even after such optimization procedure, we arrive at multiple solutions. Heuristically, some solutions are eliminated on the basis of $\{T_1,d\}$ that are too low or too high.

\section{Time Optimal Spectrum Sensing : Stochastic Optimization}
\textbf{Case A:}
$\{q_1,q_2,q_3,\ldots,q_{M}\}$ are unsorted probabilities.    
$\{p_1,p_2,p_3,\ldots,p_{M}\}$ are sorted increasing probabilities.
\begin{equation}
Mean = \hat{T} = \sum_{i=1}^{M} T_i q_i = \sum_{j \in R} \tilde{T_j} p_j = E[Z]
\end{equation}
where Z is spectrum sensing time random variable,  $\tilde{T_j}$'s are sorted sensing time values and R is a suitable index set.
\begin{equation}
\begin{split}
E[Z^2] &= \sum_{j \in R} \tilde{T_j}^2 p_j \\
Variance[Z] &= E[Z^2] - [E[Z]]^2
\end{split}
\end{equation}
\textbf{Note:} Minimizing E[Z] maximizes variance [Z]. Our goal is to minimize E[Z] as well as variance [Z] (Joint Optimization Problem). We would like to arrive at a PARETO Optimal Solution.
\\
\textbf{Note:} Suppose $q_i$'s are all equal. Then 
\begin{equation}
\begin{split}
T_i &= T_1 \text{  , for } 1 \leq i \leq M \\
E[Z] &= \sum_{j=1}^{M} \tilde{T_j} p_j = T_1
\end{split}
\end{equation}
\textbf{First Approach:}
Suppose $T_i$'s are in arithmetical progression.
\begin{equation}
\begin{split}
\text{\space}E[Z] &=\sum_{j=1}^{M} [\tilde{T_1} + (j-1)d]p_j \\
&=\tilde{T_1} (1) + \sum_{j=1}^{M}(j-1) p_j d \\
E[Z] &=\tilde{T_1} + (\mu) (d) 
\end{split}
\end{equation}
$\text{ where }   \mu = \sum_{j=1}^{M} (j-1) p_j$ 
\begin{equation}
\begin{split}
E[Z^2] &=\sum_{j=1}^{M} (\tilde{T_j})^2 p_j  \\
&=\sum_{j=1}^{M} [\tilde{T_1} + (j-1)d]^2 p_j  \\
&=\sum_{j=1}^{M} [ \tilde{T_1}^2 + (j-1)^2 d^2 + 2(j-1) \tilde{T_1} d] p_j \\
&=\tilde{T_1}^2 + d^2 \sum_{j=1}^{M}(j-1)^2 p_j + 2 \tilde{T_1}d\sum_{j=1}^{M} (j-1) p_j \\
&=\tilde{T_1}^2 + (\alpha) d^2 + (2 \tilde{T_1} d) (\mu) 
\end{split}
\end{equation}
$\textrm{ where } \alpha = \sum_{j=1}^{M}(j-1)^2 p_j$
\begin{equation}
\begin{split}
var[Z] &=\tilde{T_1}^2+(\alpha)d^2 + (2 \tilde{T_1} d)\mu -(\tilde{T_1}^2 + \mu^2 d^2 + 2\mu \tilde{T_1} d) \\
&= (\alpha) (d^2) - \mu^2 d^2 \\
&= (\alpha - \mu^2) d^2 
\end{split}
\end{equation}
\textbf{Note:}Optimal choice of $\{T_1,d\}$ are decoupled. 
Thus, the problem boils down to minimize E[Z] as well as var[Z]. How can we select the best solution?
\begin{equation}
\begin{split}
E[Z] &= \tilde{T_1} + (\mu) (d) \\
var[Z] &= (\alpha - \mu^2) d^2
\end{split}
\end{equation}
$\mbox{ where }   \mu = \sum_{j=1}^{M}(j-1)p_j      \mbox{ and }   \alpha = \sum_{j=1}^{M}(j-1)^2 p_j $
\\
i.e.$\{\mu,\alpha\}$  are determined by probabilities $\{\hat {p_j}\}_{j=1}^{M}$ and are fixed / constants.
\\
\textbf{Problem:} Determine $\tilde{T_1}$ and 'd'  from possibly non-unique solutions for $\{\tilde{T_1},d\}$ (determined by Diophantine equation)
\\
\textbf{Note:} $\tilde{T_1}$ does not effect var[Z] and only affects E[Z]. So choose minimum possible positive solution for $\tilde{T_1}$. 
\\
Simultaneously minimize E[Z], var[Z] with respect to 'd' (treating $\tilde{T_1}$ as constant.)
\begin{equation}
\begin{split}
E[Z] &=f(d)= \tilde{T_1} + (\mu) (d) \\
var[Z] &=(\alpha-\mu^2) d^2 \\
\mbox{ hence} (\alpha - \mu^2) &\geq 0
\end{split}
\end{equation}
\textbf{Note:}If only mean needs to be minimized, choose the smallest $\tilde{T_1}$ and matching value for d among pairs of solution of \eqref{eq:1}.  \\
\textbf{Note:}It can easily be reasoned that, with $\tilde{T_1}$ being chosen as smallest feasible value, d is chosen to be smallest matching value from among all solutions of Diophantine equation \ref{eq:1}.
\\
\textbf{Lemma:}Unique optimal solution for d exists where E[Z]=var[Z].\\
\textbf{Proof:}
For an optimal  solution 
\begin{equation}
\begin{split}
E[Z] &=var[Z] \\
\tilde{T_1} + (\mu) (d) &= (\alpha - \mu^2) d^2 \\
(\alpha - \mu^2) d^2 -\mu d - \tilde{T_1} &=0 \\
ad^2 + bd+c &= 0
\end{split}
\end{equation}
$\mbox{   where  } a=(\alpha - \mu^2), b= -\mu, c= - \tilde{T_1}$
\begin{equation}
\begin{split}
b^2-4ac &> 0 \text{      for d to be real.} \\
\mu^2 - 4(\alpha - \mu^2)(- \tilde{T_1}) &> 0 \\
\mu^2+ 4 (\alpha - \mu^2 )\tilde{T_1} &> 0 \mbox{ since } (\alpha - \mu^2) > 0
\end{split}
\end{equation}
\\
The zeros are distinct, thus we are interested in the value of 'd' in the first quadrant. Thus,a unique optimal solution for 'd' is achieved. Q.E.D.
\\
\textbf{Note:} We expect the optimization problem formulated in the time-optimal spectrum sensing to arise in other applications. The above lemma provides solution.\\
\\
\textbf{Case B:}
Suppose $T_i$'s are in geometrical progression.
\begin{equation}
\begin{split}
E[Z]&=\sum_{j=1}^{M} \tilde{T_j} p_j \\
E[Z]&=\sum_{j=1}^{M} (\tilde{T_1} d^{j-1})p_j \\
&= \tilde{T_1}(\sum_{j=1}^{M} d^{j-1} p_j) \\
E[Z^2]&=\sum_{j=1}^{M}(\tilde{T_j})^2 p_j \\
&=\sum_{j=1}^{M} ({\tilde{T_1}}^2 d^{2(j-1)})p_j \\
var[Z]&=E[Z^2] - (E[Z])^2 \\
&= {\tilde{T_1}}^2 [\sum_{j=1}^{M} d^{2j-2} p_j] - {\tilde{T_1}}^2 [\sum_{j=1}^  {M} d^j p_j]^2 \\
&= {\tilde{T_1}}^2[ (\sum_{j=1}^{M} d^{2j-2} p_j)-(\sum_{j=1}^{M} d^j p_j)^2]
\end{split}
\end{equation}
\textbf{Note:}
\begin{equation}
\begin{split}
E[Z] &=\tilde{T_1} f(d) \\
var[z] &= {\tilde{T_1}}^2 R(d) 
\end{split}
\end{equation}
\begin{equation}
\begin{split}
\text{       where } f(d) &= \sum_{j=1}^{M} d^j p_j \text{  and } \nonumber \\
R(d) &= (\sum_{j=1}^{M} d^{2j-2} p_j) - (\sum_{j=1}^{M} d^j p_j)^2 \\
&= f(d^2) - [f(d)]^2
\end{split}
\end{equation}
Thus the optimal choice of minimal $\tilde{T_1}$ will be optimal for both E[Z] and var[Z]. But minimization of E[Z] with respect to d will maximize var[z]. Thus we are interested in Pareto Optimal Solution i.e. jointly optimal choice for 'd' for minimizing E[Z] as well as var[z]. We now prove that if f(d) is minimized, R(d) is maximized.\\
\textbf{Claim:} If f(t) is minimized, then $f(t^2)$ is maximized, which leads to Pareto optimal solution. \\
Suppose we consider the unconstrained optimization/minimization of f(t) then
\begin{equation}
\begin{split}
\mbox{Let } K(t) = t^2 \Rightarrow f(t^2) = f(K(t))\\
\frac{df(K(t))}{dt} = \frac{df}{dk} \frac{dk}{dt} = (\frac{df}{dt})(2t)\\
\frac{d^2f(K(t))}{dt^2} = \frac{d^2f}{dt^2}(2t) + (\frac{df}{dt})(2)\\
\frac{df(K(t))}{dt} = 0 \mbox{ if and only if } \frac{df}{dt} = 0
\end{split}
\end{equation}
Further, the minima of f(.) are in the left half plane. Suppose they are real valued, e.g. to 
\begin{equation}
\begin{split}
\frac{d^2f}{dt^2} \vert_{t=t_0} > 0 \mbox{     with $t_0 < 0$}\\
\frac{d^2f(K(t))}{dt^2}\vert_{t=t_0} = \frac{d^2f}{dt^2} (2t)\vert_{t=t_0} + 0\\
\mbox{ Since  }t < t_0 \\
\frac{d^2f(K(t))}{dt^2}\vert_{t=t_0} < 0
\end{split}
\end{equation}
$f(t^2)$ is maximized, when f(t) is minimized. Thus, we look for Pareto optimal solution for d i.e. denoted t here. So use closest solution of 7 pair of ${T_1, d}$ to Pareto optimal solution.
\\\\
\textbf{Pareto Optimal Solution:}\\
\textbf{Fixed Point Equation:}
\begin{equation}
\begin{split}
E[Z]&=\tilde{T_1} f(d) \\
E[Z^2]&=\tilde{T_1}^2 g(d)\\
E[Z] &= var[Z] \\
\mbox{  note that  g(d) = f}(d^2) \\
\tilde{T_1} f(d) &= \tilde{T_1}^2 g(d)- \tilde{T_1}^2 (f(d))^2 \\
\tilde{T_1} f(d^2)  - \tilde{T_1} (f(d))^2 - f(d) &= 0
\end{split}
\end{equation}
Since f(d) is a polynomial in d, we have a polynomial equation which has multiple zeros.\\
\textbf{Q:} How can we determine optimal 'd'?\\
Choose smallest real 'd' that is feasible.\\
\textbf{Example:} 
Let M=3
\begin{equation}
\begin{split}
f(d) &= \sum_{j=1}^{3} d^{(j-1)} p_j = p_1 + d p_2+d^2 p_3 \\
g(d) &= f(d^2) = \sum_{j=1}^{3} d^{2j-2} p_j = p_1 + d^2 p_2+d^4 p_3 
\end{split}
\end{equation}
Replace values in equation
\begin{equation}
\tilde{T_1} f(d^2) - \tilde{T_1} (f(d))^2 - f(d) = 0 
\end{equation}
Let $\tilde{T_1} = 1, p_1 = .5, p_2 = .3, p_3 = .2 $
$$(p_1 + d^2 p_2+d^4 p_3) - (p_1 + d p_2+d^2 p_3)^2 - (p_1 + d p_2+d^2 p_3) = 0  $$
After solving equations
values for d =
2.43, -1.26, -0.20$\pm$0.68i \\
\\
Let $\tilde{T_1} = 1, p_1 = .2, p_2 = .3, p_3 = .5$\\ 
values for d =
2.53, -1.19, -0.06$\pm$0.21i\\
\\
Let $\tilde{T_1} = 1, p_1 = .6, p_2 = .3, p_3 = .1 $\\
values for d =
2.69, -1.33, -0.34$\pm$0.99i\\
\\
In the above examples, the solution contains only one positive real value that is of our interest. Rest of the values are not useful. Take the closest integer value of d which is real positive optimal solution.\\\\
\textbf{Summary:}\\
\textbf{Step1:} Based on data, predict the Probabilities related to spectrum band occupancy.\\
\textbf{Step2:} Allocate “Sensing Times” in the order of probability values i.e. if a band is highly occupied (probabilistic), allocate smaller sensing time and vice-versa.\\
\textbf{Step3:} Assume that the sensing times are in arithmetic/ Geometrical progression. Compute solution to the Integer programming problem (or the Diophantine equation.)
If there is more than one solution, we need to decide the solution that must be chosen.\\
\textbf{Step4:} Find solution/solutions which minimize the mean, variance (assuming that the sensing time values are in AP/GP) and find unique/multiple solutions.\\
In summary if the allocated times are in AP:
\begin{enumerate}
\item
If E[Z] and Var[Z] both require minimization, choose smallest $\{T_1,d\}$ pair solution to \eqref{eq:1}.
\item
If E[Z] is maximized and Var[Z] require minimization, it will lead to unique Pareto optimal solution.
\item
If Var[Z] is maximized and E[Z] require minimization, it will lead to unique Pareto optimal solution.
\end{enumerate}
if the allocated times are in GP:
\begin{enumerate}
\item
If E[Z] and Var[Z] both require minimization, choose Pareto solution to d rounded \& closest matching pair $\{T_1, d\}$ solution to \eqref{eq:3}, with ${T_1}$ chosen as small as possible.
\item
If E[Z] is maximized and var[Z] require minimization, choose Pareto solution to d rounded \& closest matching pair $\{T_1, d\}$ solution to \eqref{eq:3}, with ${T_1}$ chosen as large as possible..
\end{enumerate}
\textbf{Case C:} Suppose $T_i$'s are in Arithmetico-geometric progression.
\begin{equation} \label{eq:ap_gp_1}
\begin{split}
E[Z]&=\sum_{j=1}^{M} \tilde{T_j} p_j \\
E[Z]&=\sum_{j=1}^{M} [\tilde{T_1} +(j-1)d]r^{j-1} p_j \\
&= \tilde{T_1}\sum_{j=1}^{M} r^{j-1} p_j + d \sum_{j=1}^{M} (j-1) r^{j-1} p_j\\
&= \tilde{T_1} f_1(r)+d f_2(r)\\
\mbox{where } f_1(r) &= \sum_{j=1}^{M} r^{j-1} p_j \mbox{ and } f_2(r) = \sum_{j=1}^{M} (j-1) r^{j-1} p_j \\
E[Z^2] &=\sum_{j=1}^{M}(\tilde{T_j})^2 p_j \\
&=\tilde{T_1}^2\sum_{j=1}^{M} r^{2(j-1)} p_j +  d^2 \sum_{j=1}^{M} [(j-1) r^{j-1}]^2 p_j+ \\& 2T_1 d\sum_{j=1}^{M}(j-1)r^{2(j-1)}p_j \\
&= \tilde{T_1}^2 f_3(r) + d^2 f_4(r) + 2T_1 d f_5(r) \\
\mbox{where } f_3(r) &= \sum_{j=1}^{M} r^{2(j-1)} p_j \mbox{ , } f_4(r) = \sum_{j=1}^{M} [(j-1) r^{j-1}]^2 p_j\\
\mbox{and } f_5(r) &= \sum_{j=1}^{M} (j-1) r^{2(j-1)} p_j\\
\end{split}
\end{equation}
\begin{equation}\label{eq:ap_gp_2}
\begin{split}
\mbox{Case: If r = d}\\
E[Z] &= \tilde{T_1}\sum_{j=1}^{M} d^{j-1} p_j +  \sum_{j=1}^{M} (j-1) d^{j} p_j\\
&= \tilde{T_1} f_1(d)+d f_2(d)\\
\mbox{where } f_1(d) &= \sum_{j=1}^{M} d^{j-1} p_j \mbox{ and } f_2(d) = \sum_{j=1}^{M} (j-1) d^{j} p_j\\
E[Z^2]&=\tilde{T_1}^2\sum_{j=1}^{M} d^{2(j-1)} p_j +  \sum_{j=1}^{M} [(j-1) d^{j}]^2 p_j+ \\& 2T_1 \sum_{j=1}^{M}(j-1)r^{2j-1}p_j \\
&= \tilde{T_1}^2 f_3(d) + f_4(r) + 2T_1 f_5(r) \\
\mbox{where } f_3(d) &= \sum_{j=1}^{M} d^{2(j-1)} p_j \mbox{ , } f_4(d) = \sum_{j=1}^{M} [(j-1) d^{j}]^2 p_j\\
\mbox{and } f_5(d) &= \sum_{j=1}^{M} (j-1) r^{2j-1} p_j\\
\mbox{Variance }	 var[Z]&=E[Z^2] - (E[Z])^2 
\end{split}
\end{equation}

For a Pareto Optimal Solution 
\begin{equation}
\begin{split}
E[Z] &= var[Z]\\
E[Z] &= E[Z^2] - {E[Z]}^2\\
\end{split}
\end{equation}
Keep values from equation \eqref{eq:ap_gp_1} and \eqref{eq:ap_gp_2} and solve the functional equation in $\{T_1,d\}$.\\

TODO: Numerical Experiments
\\\\
\textbf{Case D:}
\textbf{Generalization:}\\
Sensing times form an increasing sequence (not necessarily AP/GP). 
$\tilde{T_1},\tilde{T_2},\ldots,\tilde {T_M}\mbox{        are such that} $
\begin{equation}
\tilde{T_1}+\tilde{T_2}+\ldots+\tilde{ T_M} = L
\end{equation}
 $$\mbox{   where     }\tilde{T_i} > 0 \mbox{ for } 1 \leq i \leq M$$
We have a constrained partition problem (as in Number Theory.) i.e. Find all possible solutions of partition problem and prune out unsuitable solutions based on some criterion. $ \tilde{T_i} < \tilde{T_j} \mbox{  if  } j > i $\\
With this constraint only, the number of possible solution need to be computed.\\
\textbf{Case 1:} $M < L$ (Most interesting case)
\begin{equation}
L (L-1) \ldots (L-(M-1)) = \frac{L!}  {(L-M+2)!}
\end{equation}
possible solutions when there is no further constraint on values of .
We don't worry about other case $M>L$.\\
Note: We can have a lower bound of sensing time allocated in any of the sub-bands i.e. $\tilde T_i > s \mbox{   for  } 1 \leq i \leq M$\\
Max number of solutions
\begin{equation}
\begin{split}
&=(L-s) (L-s-1)\ldots(L-s-(M-1)) \\
&=(L-s) (L-s-1)\ldots(L-s-M+1) \\
&=\frac{(L-s)!} {(L-s-M+2)!} 
\end{split}
\end{equation}
\textbf{Effective Idea:}\\
The most general choice of sensing times (increasing numbers) leads to the constrained partition problem. Further the sensing times must minimize the mean as well as variance of the sensing time random variable. \\
The above discussion naturally leads to the following more interesting optimization problems (related to joint optimization of moments of a discrete random variable.) 
Let 'Z' be a random variable assuming values $\{T_1,T_2,\ldots,T_M\}$ with probabilities $\{q_1,q_2,\ldots,q_M\}$ respectively. \\
\begin{equation}
E[Z^2] = \sum_{i=1}^{M} T_i ^2 q_i
\end{equation}
Let $\{T_i\}_{i=1}^M$ be the unknowns and $\{q_i\}_{i=1}^M$ are known constants. Then the “mean” and “variance” of the random variable are given by
\begin{equation}
\begin{split}
E[Z] &=\sum_{i=1}^{M} T_i q_i = f(T_1, T_2,\ldots,T_M) = f(\bar{T}) \\
var[Z] &= E[Z^2] - (E[Z])^2\\
&= g(T_1,T_2,\ldots,T_M)=g(\bar{T})
\end{split}
\end{equation}
\textbf{Goal:} To see if we can optimize E[Z], var[z] jointly.\\
\\
\textbf{Q:} Do we have an interesting functional equation arising in the joint optimization of E[Z], var[Z] ?
\begin{equation}
\begin{split}
E[Z]&=var[Z] \\
E[Z]&=E[Z^2]-(E[Z])^2 \\
E[Z^2]-E[Z]-(E[Z])^2&=0 \\
\mbox{ letting }E[Z^2] &= h(T_1, T_2,\ldots,T_M) \\
&= f(T_1 ^2, T_2 ^2,\ldots, T_M ^2)
\end{split}
\end{equation}
The multivariate functional equation that must be solved is given by 
\begin{equation}
\begin{split}
f(T_1 ^2, T_2^2,\ldots,T_M ^2) - f(T_1,T_2,\ldots,T_M)\\  - (f(T_1,T_2,\ldots,T_M))^2 = 0
\end{split}
\end{equation}
Is there a solution to such a functional equation? Mostly it constitutes the Pareto Front(Non-Dominating solution set).

\section{Multi-objective Optimization: Linear and Quadratic Programming (Hybrid Programming)}

\textbf{Objective Functions:} 
\begin{equation}
\begin{split}
& C = [ {p_1}, {p_2}, \ldots, {p_M}]^T\\
& T = [ T_1, T_2,\ldots, T_M]^T \\
& D = diag\{{p_1}, {p_2}, \ldots, {p_M}\} \\
& E[Z] = {C^T} T = {T^T} C = C.T \\
& Var[Z] = {T^T}\bar{D}T - ({C^T} T)^2 = {T^T}\bar{D}T - ({T^T} C)^2 \\
&= {T^T}\bar{D}T - {T^T} C{C^T}T = {T^T}(\bar{D}-{C^T} C)T \\
&= {T^T}\bar{G}T \mbox{ ,where } \bar{G} = \bar{D} - C{C^T}
\end{split}
\end{equation}
Example:
\begin{equation}
\begin{split}
& \bar{G} = \bar{D} - C {C^T} \\
& \bar{G} =
\begin{bmatrix}
p_1 & 0   & 0 \\
0   & p_2 & 0 \\
0   & 0   & p_3
\end{bmatrix} -
\begin{bmatrix}
{p_1}^2 & p_1 p_2   & p_1 p_3 \\
p_1 p_2   & {p_2}^2 & p_2 p_3 \\
p_1 p_3   & p_2 p_3   & {p_3}^2
\end{bmatrix} 
\\
&= \begin{bmatrix}
{p_1}(1-{p_1}) & -p_1 p_2   & -p_1 p_3 \\
-p_1 p_2   & {p_2}(1-{p_2}) & -p_2 p_3 \\
-p_1 p_3   & -p_2 p_3   & {p_3}(1-{p_3})
\end{bmatrix}
\end{split}
\end{equation}
Inferences:
\begin{itemize}
\item $\bar{G}$ is a laplacian like matrix.
\item -$\bar{G}$ is a symmetric generator matrix.
\end{itemize}

Function of interest for arriving at solutions where E[Z]=var[Z]:
\begin{equation}
\begin{split}
& J(T) = Var[Z] - E[Z] \\
&= {T^T}\bar{G}T - {T^T}C\\
&= 
\begin{bmatrix}
T^T & 1
\end{bmatrix}
\begin{bmatrix}
\bar{G} & 0 \\
-C^T & 0
\end{bmatrix}
\begin{bmatrix}
T \\
1
\end{bmatrix}
\end{split}
\end{equation}
\\
\textbf{Note:} $\bar{G} \bar{e} = \bar{0}$\\
$\bar{G} = \bar{G}^T$\\
\textbf{Note:}We use Laplacian and Laplacian like matrix interchangeably.\\
\textbf{Theme:} Laplacian matrix arising in variance optimization of a discrete random variable.\\
\textbf{Q:} Can (linear algebraic) properties of matrix G be capitalized to derive new results on variance minimization? 
\\
\textbf{Goal:} To study properties of laplacian like matrix $\bar{G} = \bar{D} - \bar{C} C^T$ where $\bar{D} = diag\{p_1, p_2,\ldots,p_M\}$, $\bar{C} = [p_1, p_2, \ldots, p_M]^T$, $\bar{G}=\bar{G}^T$
\begin{enumerate}
\item Eigen values are all real.
\item $\bar{G}$ is positive semidefinite, with an eigen value at zero ($\bar{e}\ldots$ all ones vector). $\bar{e}$ is in the null space of $\bar{G}$.
\item 0 is the smallest eigen value and all other eigen values lie on real axis.
\item Bounds on spectral radius of $\bar{G}$
\begin{equation}
\begin{split}
\sum_{j=1}^n G_{ij} &= \sum_{j=1}^n D_{ij} -p_j\sum (p_1+\ldots+p_{j-1}+p_j+\ldots+p_M)\\
&= p_j - p_j = 0 \\
\sum_{j=1}^n  |G_{ij}| &= p_j(1-p_j) + |p_j(p_1+\ldots+p_{j-1}+p_{j+1}+\ldots+p_M)| \\
&= p_j(1-p_j)+|p_j(1-p_j)| \\
&= 2p_j(1-p_j)\\
\min_{j} & \{2p_j (1-p_j)\} \leq \mbox{Spectral radius(G)} \leq \max_{j} \{2p_j (1-p_j)\}
\end{split}
\end{equation}
\end{enumerate}
All eigen values of G lie in the interval [0,1).\\
\\
\textbf{Note:}
\begin{equation}
\begin{split}
& \hat{G} = -\bar{G} \mbox{  is a generator matrix.}\\
& \hat{G}/\theta + I = P \mbox{  , stochastic matrix.}\\
& I - G/\theta = P \\
& \mu\ldots \mbox{  eigen value of }\hat{G}.\\
& \lambda\ldots \mbox{  eigen value of P}.\\
& \theta\ldots \mbox{  largest diagonal element of }\hat{G}.\\
& \epsilon\ldots \mbox{  eigen value of G}.\\
& \epsilon_0\ldots Sp(G).\\
& \mu/\theta + 1 = \lambda   \mbox{   and    } \mu = -\epsilon\\
& \mu = 0 \mbox{  , so  } \lambda = 1 
\end{split}
\end{equation}
Thus, by Perron Frobenius theorem, the dimension of null space of $\bar{G}$ is one (with $\bar{e} = \begin{bmatrix}
1 & 1 & \ldots & 1
\end{bmatrix}^T$) i.e. all ones column vector using null space.
\\\\
Computation of determinant and trace of G:
\begin{equation}
\begin{split}
& \bar{G} \bar{C} = \bar{D}\bar{C} - \bar{C} (\sum_{j=1}^M {p_i}^2)\\
&= \begin{bmatrix}
{p_1}^2 \\{p_2}^2 \\\vdots\\{p_M}^2 
\end{bmatrix}
- \delta \begin{bmatrix}
p_1 \\p_2 \\\vdots\\p_M
\end{bmatrix} \\
&= \begin{bmatrix}
{p_1}^2 - \delta p_1\\
{p_2}^2 - \delta p_2\\
\vdots\\
{p_M}^2 - \delta p_M
\end{bmatrix}\\
& \bar{G} = \bar{D} - \bar{C}\bar{C}^T \\
&= \bar{D}[I - \bar{D}^{-1} \bar{C} \bar{C}^T ]\\
& Det(\bar{G}) = Det(\bar{D}) Det[I - \bar{D}^{-1} \bar{C} \bar{C}^T ]\\ \mbox {Note:   } Det(\bar{G}) = 0\\
& \bar{D}^{-1} \bar{C} \bar{C}^T = \begin{bmatrix}
1/p_1 & 0 &\ldots & 0\\
0 & 1/p_2 &\ldots & 0\\
\vdots & \vdots & \vdots & \vdots \\
0 & 0 & \ldots & 1/p_M
\end{bmatrix}\\
& \bar{C} = \begin{bmatrix}
p_1 \\ p_2 \\ \vdots \\ p_M \end{bmatrix}; {\bar{D}}^{-1} \bar{C} = \begin{bmatrix}
p_1/p_1 \\ p_2/p_2 \\ \vdots \\ p_M/p_M  
\end{bmatrix} = \begin{bmatrix}
1 \\ 1\\ \vdots \\1
\end{bmatrix}\\
\bar{D}^{-1} \bar{C} \bar{C}^T &= \begin{bmatrix}
1 \\ 1\\ \vdots \\1
\end{bmatrix} \begin{bmatrix}
p_1 & p_2 &\ldots& p_M 
\end{bmatrix} = F\\
\end{split}
\end{equation}
It is a rank one matrix.\\
From Kailath ("Linear Systems"), page 658, we have that if A is a rank one matrix $Det(I+A) = 1+trace(A)$\\
\textbf{Determinant:}
\begin{equation}
\begin{split}
& Det(\bar(I) + (-F)) = 1+trace(-F)\\
& = 1- (p_1+p_2+\ldots+p_M) = 0\\
& Det(G) = Det(D) Det(I- D^{-1}\bar(C)\bar{C}^T) = \prod_{i=1}^N \lambda_i\\
& \prod_{i=1}^N \lambda_i = \begin{pmatrix}
& p_1 & p_2 & \ldots & p_M
\end{pmatrix} (1-p_1-p_2\ldots-p_M) = 0\\
\end{split}
\end{equation}
\textbf{Trace:}
\begin{equation}
\begin{split}
& Trace(G) = Trace(D) - Trace(\bar{C} \bar{C}^T)\\
& \sum_{i=1}^N \lambda_i = (p_1+p_2\ldots+p_M)-(p_1^2+p_2^2\ldots+p_M^2) \\
&= \sum_{i=1}^N \lambda_i > 0
\end{split}
\end{equation}

\textbf{Properties of laplacian type matrix arising in variance expression of a Discrete Random Variable Z:}\\
$Var[Z]=\bar{T}^T G \bar{T} \geq 0$\\
\begin{enumerate}
\item Global minimum occurs at the eigen vector(s) of G corresponding to zero eigen value(null space of G).
\item Let $\mu_1 \geq \mu_2 \geq \ldots \leq \mu_{N-1} \leq \mu_N = 0$. Non-zero minimum value of Var[Z] is determined by the eigen vector $\bar{R}$, corresponding to second smallest eigen value $\mu_{N-1}$.
\begin{equation}
\begin{split}
& G \bar{R} = \mu_{N-1} \bar{R} \\
& \bar{R}^T \bar{R} = \mu_{N-1} \bar{R}^T \bar{R} \\
&= \mu_{N-1} (||\bar{R}||)^2 \mbox{  ,where $||\bar{R}||$ is the $L^2$-norm of $\bar{G}$} 
\end{split}
\end{equation}
\textbf{Note:}Optimization of Var[Z] requires specification of constraint set on $\bar{T}$.\\
\textbf{Example:}
\begin{equation}
||\bar{T}||=1 \Rightarrow
\bar{R}^T \bar{G} \bar{R} = \mu_{N-1} > 0
\end{equation}
\item Using Rayleigh's theorem, when constraint set is euclidean hyper sphere $\mu_1$ is the maximum value.
\item Similar results are derived when the constraint set is unit hypercube, lattice.
\end{enumerate}
\textbf{Results related to Laplacian like matrix G:}\\
$\bar{G} = \bar{D} - \bar{C} \bar{C}^T$ is symmetric laplacian like matrix.
\begin{enumerate}
\item 
\begin{equation}
\begin{split}
& Trace(\bar{G}) = \sum_{i=1}^N(p_i - {p_i}^2)\\
&= \sum_{i=1}^N p_i - \sum_{i=1}^N {p_i}^2 \\
&= 1 - \sum_{i=1}^N {p_i}^2\\
&= 1 - \mbox{ Tsallis entropy of ($\bar{P} = (p_1,p_2,\ldots,p_N)$)}\\
&\mbox{Tsallis entropy and Shannon entropy are related.}\\
Trace(\bar{G})&=  1 - \sum_{i=1}^N {p_i}^2 \geq 0 (\mbox{ Note that   }\mu_i \in (0,1])\\
&= \sum_{i=1}^{N-1} \mu_i\\
\end{split}
\end{equation}
\item Det(G) = 0 , other coefficients of characteristic polynomial may easily be computed.
\item Computation of eigen values of G\\
$G = \sum_{i=1}^{N-1} \mu_i \bar{f_i} \bar{f_i}^T$, where $f_i$'s are the right eigen vectors of $\bar{G}$.\\
\textbf{Q:}How do we compute $f_i$'s.
\item G is sub-stochastic since $\mu_i \in [0,1)$\\
$G^n \underrightarrow{n \uparrow \infty} 0 \mbox{        ,where      } G=\bar{D} - \bar{C} \bar{C}^T$\\
Using matrix binomial theorem, $G^n$ can be explicitly computed. 
\begin{equation}
\begin{split}
& \bar{G} = \bar{D} - \hat{C} \mbox{     ,where $\hat{C}$ is rank one matrix. }\\
& \hat{C}^2 = \bar{C} \bar{C}^T \bar{C} \bar{C}^T 
= (\bar{C}^T \bar{C}) \bar{C} \bar{C}^T \\
&=  (\sum_{i=1}^N {p_i}^2) (\bar{C} \bar{C}^T ) = \alpha (\bar{C} \bar{C}^T )\\
& \hat{C}^m \mbox{ can be computed for m }\geq 2 \\
& \hat{C}^m = (\alpha)^{m-1} (\bar{C} \bar{C}^T ) \\
Also D^m &= diag\{ {p_1}^m, {p_2}^m,\ldots,{p_N}^m\}\\
& \mbox{Using expression for $G^M$, we can compute}\\
Trace(G^m) &= \sum_{i=1}^{N-1} (\mu_i)^m \mbox{    ,for } m \geq 1
\end{split}
\end{equation}
\item Using Leverrier - Fadeev algorith, all the coefficients of characteristic polynomial of G could be computed efficiently. They involve $\{p_1, p_2, \ldots, p_N \}$.
\end{enumerate}

\section{Future Work}
Consider the packet arrivals to each secondary user constitute a Poisson process. Let these packet streams be independent. Also, let there be 'K' channels available for communication. Let the service times (for transmitting the packets from the secondary users) be exponential random variables. 
Thus, we model the associated Queuing system to be an M/M/K queue. Using standard results in queuing theory, various performance measures can be computed and interpreted.
\\
\textbf{More General Stochastic Model: }
By associating channel states, a more general model based on Quasi Birth and Death process is being developed and analyzed.

\section{Conclusion}
In the paper, information theoretic and integer linear programming approach for time  optimal spectrum sensing is discussed. The problem is also discussed as stochastic optimization problem and how Pareto Front helps solving the issue. We expect the optimization problem formulated here can arise in other applications.


%



\ifCLASSOPTIONcaptionsoff
  \newpage
\fi

\end{document}